# Nonlinear field dependence of Hall effect and high-mobility multi-carrier transport in an altermagnet CrSb

**Yuqing Bai,[1] Xinji Xiang,[1] Shuang Pan,[1] Shichao Zhang,[1] Haifeng Chen,[2] Xi Chen,[1] Zhida Han,[2] Guizhou Xu,[1,*] Feng Xu[1,**]**

*1 School of Materials Science and Engineering, Nanjing University of Science and Technology, Nanjing 210094, China*

*2 School of Electronic and Information Engineering, Changshu Institute of Technology, Changshu 215500, China*

## Abstract

As a promising candidate for altermagnet, CrSb possesses a distinctive compensated spin split band structure that could bring groundbreaking concepts to the field of spintronics. In this work, we have grown high-quality CrSb single crystals and comprehensively investigated their electronic and magneto-transport properties. We have observed large, positive, and non-saturated magnetoresistance (MR) in CrSb, which well obeys Kohler's rule, indicating its classic Lorentz scattering origins. Remarkably, a nonlinear magnetic field dependence of Hall effect resembling the spontaneous anomalous Hall is identified over a wide temperature range. After careful analysis of the transport data, we conclude the non-linearity mainly stems from the incorporation of different carriers in the magnetoconductivity. According to the Fermi surface analyses of CrSb, we applied the three-carrier model to fit the conductivity data, yielding good agreement. The extracted carrier concentration and mobility indicates that CrSb behaves more like a semimetal, with the highest mobility reaching $3\times10^3$ cm$^2$/Vs. Furthermore, calculations using the semiclassical Boltzmann transport theory have successfully reproduced the main features of the experimental MR and Hall effect in CrSb. These exceptional transport properties make CrSb unique for applications in spintronics as an altermagnet.

[*] Corresponding author

[**] Corresponding author

E-mail addresses: gzxu@njust.edu.cn (G.Z. Xu), xufeng@njust.edu.cn (F. Xu)

## 1. Introduction

A novel magnetic phase, referred to as altermagnet (AM), has recently emerged from the vast family of collinear antiferromagnets and has garnered significant attention [1–3]. The AM features a compensated antiparallel spin configuration similar to that of an antiferromagnet (AFM), but with a spin split in momentum space comparable to a ferromagnet (FM). To break the global Kramers spin-degeneracy of collinear AFM, it requires that the opposite-spin sublattices are connected by a time-reversal ($\mathcal{T}$) symmetry combined with a spatial symmetry excluding real-space inversion ($\mathcal{P}$) or lattice translation ($\boldsymbol{t}$), as formalized by the spin group theory [1,2,4]. This unique $\mathcal{T}$ symmetry breaking is imposed by a local anisotropic electric crystal field resulting from the presence of asymmetric non-magnetic atoms. As a result, the spin splitting observed in AM is in stark contrast to the conventional mechanisms of ferromagnetic splitting caused by net magnetization, or relativistic spin-orbit coupling band splitting due to the absence of an inversion center. Rather, it is a non-relativistic effect stemming from the interplay of exchange interaction and crystal field effect. This can give rise to a much larger band splitting compared to the conventional Rashba effect. In this sense, AM introduces a new avenue for efficient spin-to-charge conversion and spin torque effects [5–8]. The specific spin split in AM enables the existence of non-vanishing Berry curvature, acting like a ferromagnetic dipole and giving rise to a substantial spontaneous anomalous Hall effect that previously only ascribed to FM or non-collinear AFM [9–12]. In the meantime, AM embraces the advantages of AFM, such as ultra-fast spin dynamics and less susceptibility to external magnetic fields [13,14]. These combinations suggest that AM can hopefully lead to groundbreaking concepts and applications in the realm of spintronics, magneto-optics [15], and beyond.

Based on symmetry analyses and first principles studies, several potential candidates have been predicted to exhibit altermagnetic properties[1,16]. Among these, the rutile $RuO_2$ and semiconductor MnTe in hexagonal structure have emerged as outstanding prototypes and been extensively studied [7,8,11,12,17–19]. They have demonstrated spin-split features, magnetic transport behavior, and spin torque effects that align with the exceptional properties predicted for AMs. The hexagonal NiAs-type compound CrSb (space group: $P6_3/mmc$) has also emerged as an AM candidate [1,16]. In this compound, the two antiparallel magnetic sublattices are connected by a sixfold anti-unitary rotation ($C_6\mathcal{T}$) and a half-unit cell translation (Fig. 1(a)), fulfilling the symmetry requirements of AM. Very recently, the band splitting has been confirmed in CrSb thin films by spin-integrated soft X-ray angular-resolved photoelectron spectroscopy [20]. With a high Néel temperature of approximately 700 K [21,22] and a substantial spin split magnitude of around 1 eV [1,16], metallic CrSb shows great potential as an AM material that deserves to be thoroughly studied.

In this work, we have carried out systematic magneto-transport studies on high-quality CrSb single crystals. A nonlinear magnetic field dependence of Hall effect that resembles the spontaneous anomalous Hall is identified at temperatures below 200 K. Upon careful analysis of the transport data, we determined that the non-linearity stems from the involvement of different carriers in the transverse magnetoconductivities. By considering the calculated Fermi surfaces (FSs) of CrSb, we identified three main types of carriers that dominate the transport properties. The nonlinear field dependences of Hall effect and magnetoresistances are well reproduced by fitting the three-carrier model within the realm of classical Drude theory. A large magnetoresistance up to 27% is observed at 10 K and 9 T, which relates to the high-mobility of the carriers reaching $10^3$ $cm^2/Vs$. These findings highlight the potential of CrSb as a promising altermagnet candidate for spintronics applications.

## 2. Experimental Section

### 2.1. Sample preparation and characterization

CrSb single crystals were grown by the chemical vapor transport (CVT) method with iodine ($I_2$) as a transport agent. Chromium and antimony powders with a molar ratio of 1:1 was weighed and mixed with 0.25 g of anhydrous iodine beads, which were sealed in a vacuumed quartz tube. The quartz tube was placed in a dual-temperature zone tube furnace, with the hot side set at 1173 K, where a total mass of 0.75 g of raw material was put, and the cold side maintained at a temperature of 973 K. This temperature gradient was sustained for 7 days, after which the tube was gradually cooled down to room temperature in the furnace. Finally, air-stable single crystals with a shiny surface and an average size of ~ 1 × 1 × 0.3 $mm^3$ were obtained.

The element composition was examined by energy-dispersive spectroscopy (EDS) in the scanning electron microscope (SEM, FEI Quanta 250F). The selected area electron diffraction (SAED) patterns and the high-resolution lattice arrangement were acquired by high-resolution transmission electron microscopy (TEM, Thermo Fisher Talos F200S G2). For the TEM observation, the bulk sample was immobilized on Mo grids and then subjected to ion-thinning treatment to obtain the flakes. The crystal structure was identified by x-ray diffraction (XRD, Bruker-AXS D8 Advance) with Cu−Kα radiation. Differential scanning calorimetry (DSC, Mettler Toledo DSC3) was used to determine the magnetic transition temperature with a temperature variation rate of 5 K $min^{-1}$. Magnetization and transport properties were measured in the superconducting quantum interference device (SQUID, Quantum Design) and physical property measurement system (PPMS, Quantum Design), respectively. For measuring the magnetoresistance (MR) and Hall resistivity, a four-probe method was applied and the electrode contacts were made of silver paste. The final MR and Hall data were symmetrized to exclude the misalignment of the electrode.

**2.2. Calculation details**

First principles calculations were performed with the projector augmented wave (PAW) method, as implemented in the Vienna ab initio simulation package (VASP) [23,24]. The exchange-correlation effect was treated with a generalized gradient approximation (GGA) function in the form of Perdew-Burke-Ernzerhof (PBE) parametrization. The static self-consistency calculations were carried out on a $k$ grid of 9 × 9 × 9, with the cutoff energy of 500 eV for the plane wave basis set. The experimental lattice constants $a = b = 4.10$ Å and $c = 5.45$ Å were used for the calculations. We constructed the maximally-localized Wannier functions (MLWFs) by projecting the Bloch states obtained from VASP onto the Cr 3$d$ and Sb 5$p$ atomic orbitals, using the package of Wannier90 [25].Then, the FS was constructed on a $k$-mesh of 50 × 50 × 50 using Wannier90 and visualized by Xcrysden [26]. The MR and Hall resistivity are calculated using the WannierTools [27] software packages, based on the Boltzmann transport approach that relies on a semiclassical model and constant relaxation time approximation [28].

## 3. Results and discussions

Based on previous studies, CrSb is a collinear antiferromagnet with parallel magnetic moments in the *ab*-plane and anti-parallel moments between interlayer Cr atoms, which preserve the magnetic compensation [21,22,29]. The magnetic moments are mainly contributed by the Cr atoms, with the effect of the non-magnetic atoms on the magnetic structure commonly neglected. However, by considering the noncentrosymmetric Sb positions, as seen in Fig. 1(a), it becomes evident that they introduce asymmetry in the spin density on sublattices with opposite spin. These Sb atoms form rotating (or mirroring) octahedra around the Cr atoms, thus the two magnetic sublattices are connected by a $C_{6z}\mathcal{T}$ and a translation $\boldsymbol{t}$ (0, 0, $c$/2), or a $M_z\mathcal{T}$ operation, defining the symmetry of the altermagnetic phase [1,16]. This kind of $\mathcal{T}$ symmetry breaking results in the lifting of spin-degeneracy at specific *k*-paths, like the ΓL (or ΓL′) paths in the Brillouin zone, while maintaining spin-degeneracy at other high-symmetry paths conventionally used for the plot of band structure, as depicted in Fig. 1(b, c). In addition, three main types of Fermi pockets are identified in the FSs (Fig. 1(c)), including one large barrel-shaped hole and two smaller volumes of scattered electrons. One electron pocket (comprising 8 segments) is distributed along the symmetrical ΓM paths and the other (comprising 12 pieces) is located along the ΓL paths.

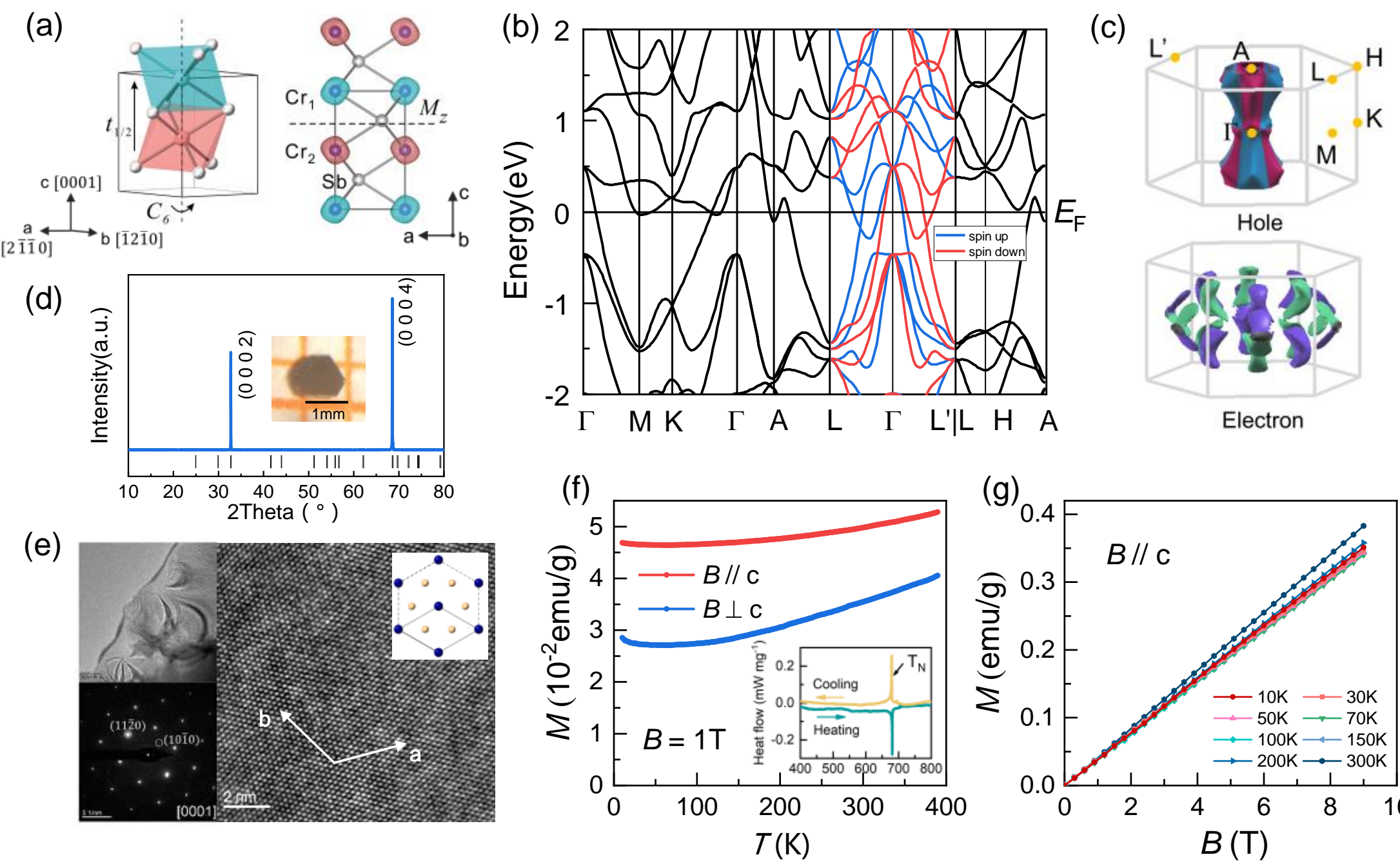

**Fig. 1.** (a) Left panel is the hexagonal NiAs-type CrSb structure, with the two magnetic sublattices with opposite spins are marked with different colors. Right panel is the calculated spin density distribution. (b) Spin split band structure of CrSb without consideration of spin-orbital coupling. The high symmetry positions in the first Brillouin zone (BZ) are Γ (0, 0, 0), M (1/2, 0, 0), K (1/3, 1/3, 0), A (0, 0, 1/2), L (1/2, 0, 1/2), L' ( -1/2, 0, 1/2) and H (1/3, 1/3, 1/2), respectively. (c) The spin-resolved Fermi surfaces. Different colors in each FS represent opposite spin direction. (d) XRD pattern and optical photograph of CrSb single crystal. (e) The edge area of the sample treated with ion thinning, the SAED patterns along the [0001] direction, and the high-resolution TEM image. Inset is the projected crystal structure. (f) The thermo-magnetization $M(T)$ curves measured at temperatures from 10 to 390 K for $B \parallel c$ and $B \perp c$. The inset shows the DSC curves measured at temperatures from 400 to 800 K. (g) The isothermal magnetization $M(H)$ curves measured at various temperatures for $B \parallel c$.

Figure 1(d) shows the optical photographs of the single crystals we grown, along with the single crystal XRD patterns. The sharp edges and quasi-hexagonal faces of the sample indicate its single crystalline nature. The XRD reveals only monotonic (00*l*) diffraction peaks belonging to the *c*-plane of the hexagonal structure. Powder XRD of crushed CrSb single crystal was also collected at room temperature (supplementary Fig. S1), further supporting the NiAs-type structure, without the presence of any impurity phases. The refined lattice constants are $a = 4.10$ Å and $c = 5.45$ Å, in good agreement with those reported [21,29]. The selective area electron diffraction (SAED) in Fig. 1(e) shows clearly-defined diffraction points exhibiting expected hexagonal symmetry. The lattice arrangement of the hexagonal structure appears sharp-edged in the high-resolution TEM image, indicating high quality of our single-crystalline sample. The lattice spacing of the $(11\bar{2}0)$ crystal plane, which exhibited the brightest spots, was also deduced, yielding $d_{(11\bar{2}0)} \approx 2.05$ Å, showing excellent agreement with the XRD results.

The thermo-magnetization $M(T)$ behaviors of CrSb single crystals were examined under a constant field parallel ($B \parallel c$) and perpendicular ($B \perp c$) to the $c$-axis, as shown in Fig. 1(f). Since the Néel temperature is well above the room temperature, the Néel transition cannot be detected in the present temperature range from 10 to 390 K. Still, we observed that the magnetization decreases with lowering temperature, implying the typical behavior of AFM. The isothermal magnetization $M(H)$ curves at various temperatures between 10 and 300 K for $B \parallel c$ and $B \perp c$ can be seen in Fig. 1(e) and supplementary Fig. S2. These curves displayed a very linear relationship, implying a strong antiferromagnetic order in CrSb without any canted component. The magnitude of $M$ is larger in $B \parallel c$, consistent with the $M(T)$ results. Since the high-temperature $M(T)$ measurements were not available, possible enthalpy change owing to the antiferromagnetic transition was detected using DSC. As shown in the inset of Fig. 1(f), large exothermic and endothermic peaks around 680 K are observed during cooling and heating processes. These transitions exhibit no hysteresis, indicating their second order origin. The transition temperature observed here is slightly lower compared to previous research [21,22].

Figure 2 exhibits the longitudinal resistance measured under varying temperatures or external magnetic fields on CrSb crystals. The quasi-hexagonal shape of the crystals allowed us to index specific crystalline orientations in the $ab$-plane and $c$-axis, denoted as $x$, $y$, $z$ directions in the following transport measurements, as seen in the inset of Fig. 2(a). The resistivity $\rho_{xx}$ continuously increases with rising temperature (Fig. 2(a)), representing typical metallic conduction behavior. Resembling a general case, at high temperatures, $\rho_{xx}(T)$ is dominated by electron-phonon scattering, showing a linear $T$ dependence, while at lower temperatures, it is mainly influenced by electron-electron scattering, displaying a $T^2$ dependence [30] (see Supplementary Fig. S3 for fitting details). Overall, it can be effectively described by $T^{1.67}$ relation. The upper inset demonstrates that when the magnetic field is ramped to 9 T, the metallic behavior of $\rho_{xx}(T)$ is not significantly altered, highlighting the stronger influence of temperature over the magnetic field on the resistivity.

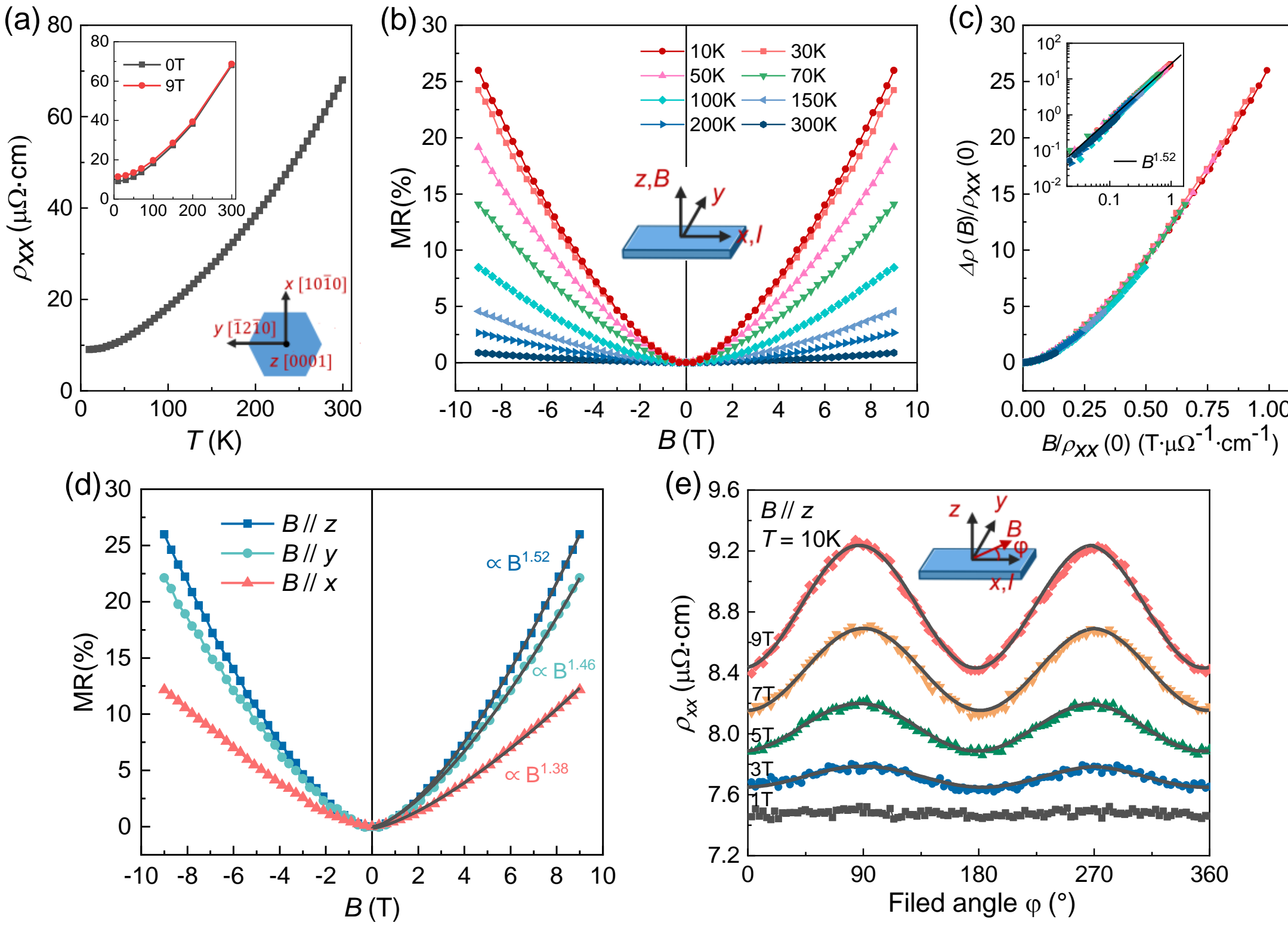

**Fig. 2.** (a) Temperature dependences of the zero-field longitudinal resistivity $\rho_{xx}$ . The upper inset is the $\rho_{xx}$ under the field of 0 and 9 T at various temperatures. The lower inset defines specific crystalline orientations in the hexagonal crystal. (b) MR at various temperatures ranging from 10 to 300 K for $B \parallel z$, $I \parallel x$. (c) $\Delta\rho_{xx}(B)/\rho_{xx}(0)$ as a function of $B/\rho_{xx}(0)$. The inset displays the same curves plotted in log coordinates, along with linear fits that reveals MR $\propto (B/\rho_0)^{1.52}$ . (d) MR at 10 K for $B \parallel x$, $y$, and $z$, $I \parallel x$. (e) Angle dependences of in-plane $\rho_{xx}$ at 10 K under the field of 1, 3, 5, 7 and 9 T. The insets in (b) and (e) show the schematics of the measurement configuration.

The isothermal magnetoresistance (MR) curves for $B \parallel c$ at various temperatures ranging from 10 to 300 K are shown in Fig. 2(b). It is noted that the MR for the collinear antiferromagnetic CrSb can reach 27% at 10 K and 9 T, surpassing the majority of commonly studied AFM metallic materials [31,32]. As temperature increasing, the magnitude of MR lowers, but remains positive across the entire magnetic field range. Following the classic Kohler's rule [33–35], all MR curves can collapse into a single one when plotted with a combined variable $B/\rho_0$. Our MR data show excellent compliance with Kohler's rule, as illustrated in Fig. 2(c), indicating that the MR is dominated by the orbital cyclotron motion of carriers under the Lorentz force, and the relaxation-time approximation works effectively in this system. Typically, when there are two kinds of nearly compensated carriers, the scaled MR would follow a quadratic behavior. Here the MR varies with a $B^{1.52}$ relation, suggesting the coexistence of different kinds of carriers with imbalanced populations [28,30].

When the magnetic field turns to the $x$ or $y$ direction, the resistivity lowers a bit, but all data follow a $B^{\alpha}$ relation at 10 K, with $\alpha$ close to 1.5 (Fig. 2(d)). The angular dependences of $\rho_{xx}$ with the magnetic field rotated in the $ab$-plane are shown in Fig. 2(e). They follow a typical cos 2φ dependence on the field-angle, mainly reflecting the difference in Lorentz force scattering when the magnetic field is parallel and perpendicular to the current [36,37]. A considerable anisotropic magnetoresistance (AMR) up to 9.8% at 10 K and 9 T is observed. It is larger than most conventional collinear AFMs [32,36], while comparable to that of the altermagnet MnTe [12,37].

The Hall effect was examined for CrSb single crystals at various temperatures ranging from 10 to 300 K. Remarkably, the $\rho_{yx}(B)$ curves exhibit significant non-linearity with the field sweeping from -9 T to 9 T, as seen in Fig. 3(a). At lower fields, the Hall coefficient ($R_H$) is negative, but with an increase in the field, it shifts towards positive. This trend becomes more pronounced at lower temperatures, where an obvious sign reversal occurs. These features of the Hall effect can be reproduced in multiple samples (supplementary Fig. S4). On one side, the non-linearity can be attributed to the spontaneous anomalous Hall effect (AHE), which has been suggested in several altermagnetic materials [11,12]. On the other side, it could also be a resulted of the multi-band carrier transport [30,38]. Since the AHE is excluded in the pristine CrSb where the Néel vector aligns along the $c$-axis [16,21] and the FSs and MR indicate multi-carrier transport, a multi-carrier model was utilized to describe the magnetoconductivity. The longitudinal conductivity $\sigma_{xx}$ and Hall conductivity $\sigma_{yx}$ are obtained by $\sigma_{xx} = \frac{\rho_{xx}}{\rho_{xx}^2+\rho_{xy}^2}$ and $\sigma_{yx} = -\frac{\rho_{yx}}{\rho_{xx}^2+\rho_{xy}^2}$, respectively. The multi-band model incorporates the contributions from various types of carriers, each characterized by carrier densities and mobilities, using the following equations [28,38]:

$$\sigma_{xx}(B) = \sum_i \frac{n_i e \mu_i}{1+u_i^2 B^2} \quad (1)$$

$$\sigma_{yx}(B) = \sum_i \frac{n_i e \mu_i^2 B}{1+\mu_i^2 B^2} \quad (2)$$

where $n_i$ and $\mu_i$ indicate the carrier concentration and mobility, respectively, with positive values indicating electrons and negative values indicating holes.

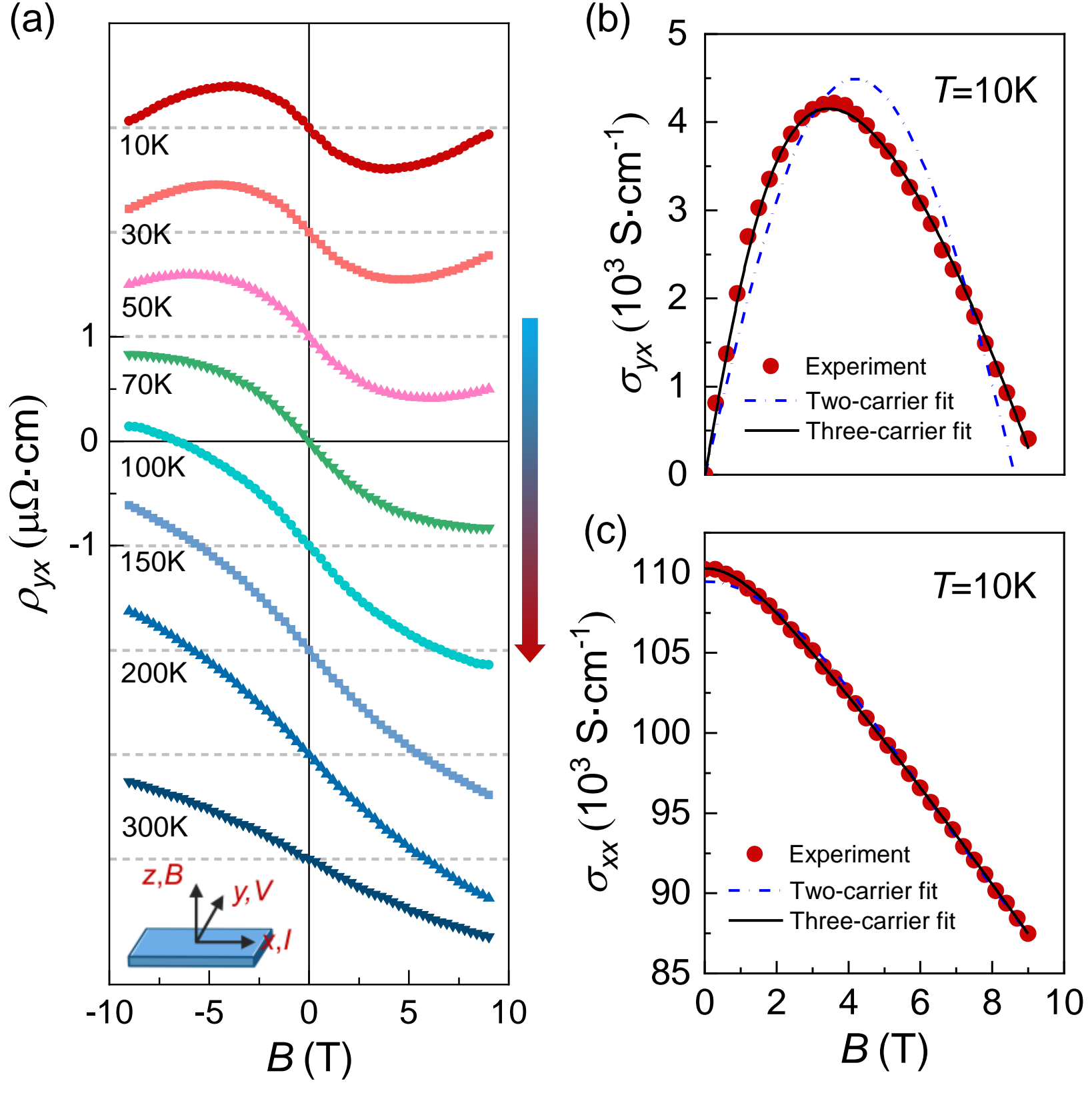

**Fig. 3.** (a) Field dependences of Hall resistivity $\rho_{yx}$ at various temperatures ranging from 10 to 300 K for $B \parallel z$. The inset is the configuration of measurement. (b, c) The Hall conductivity $\sigma_{yx}$ and longitudinal conductivity $\sigma_{xx}$ at 10 K and their fittings using the two-carrier and three-carrier models.

Initially, we considered the commonly-used two-carrier model. Our findings revealed that the data align well with the two-band model at elevated temperatures. However, at lower indicas seen in the 10K fitting (see Fig. 3(b)). The deviation in $\sigma_{xx}$ is not so severe, but still obvious at low fields (Fig. 3(c)). Given that the Fermi surfaces mainly comprise three kinds of carriers (a large hole and two electron types), we applied the three-carrier model to fit data at low temperatures. As shown in Figs. 3(b, c), both $\sigma_{yx}$ and $\sigma_{xx}$ can be well reproduced in the whole field region.

Figure 4 displays conductivity data and corresponding fittings at all measured temperatures. It can be seen that the non-linearity of the Hall conductivity is well captured by the three-carrier model. The extracted carrier concentration ($n_i$) in Fig. 4(c) reveals one larger hole pocket and two smaller electron pockets, in line with the Fermi surfaces calculations (Fig. 1(c)). The $n_i$ varies from $10^{18}$ to $10^{21}$ cm$^{-3}$ (below 100 K), with the carrier mobility ($\mu_i$) ranging from $1\times10^2$ to $3\times10^3$ cm$^2$/Vs. These values can be compared to those in semimetals or semiconductors [39,40]. The carrier with higher concentration exhibits lower mobility, leading to competition in $\sigma_{yx}(B)$ and finally causing a sign change in the observable magnetic field. At high fields, the slope of $\rho_{yx}(B)$ is determined solely by $\frac{1}{n_h-n_e}$ [30]. The positive slope of $\rho_{yx}(B)$ (Fig. 3(a)) further confirms the dominance of hole carriers.

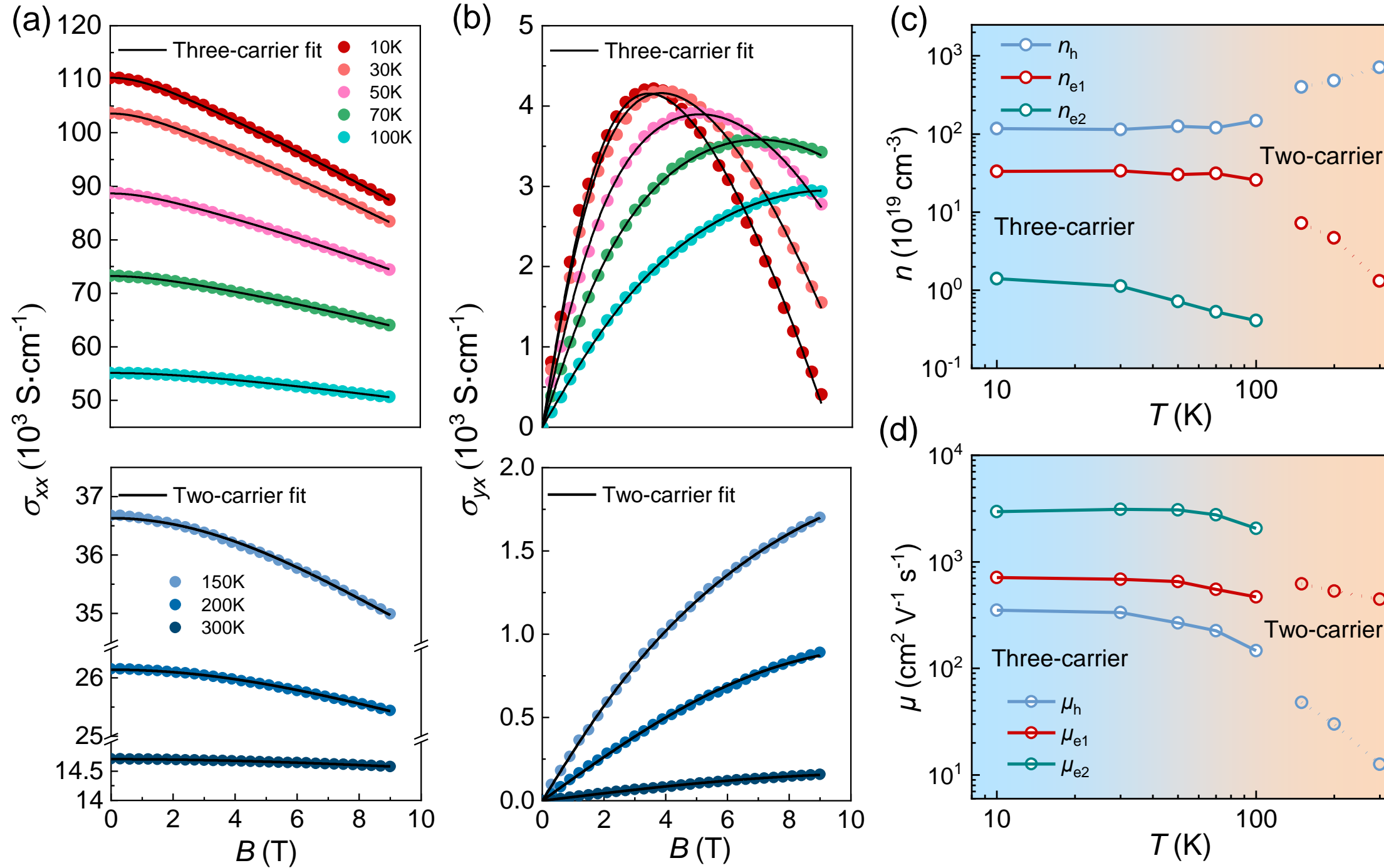

**Fig. 4.** (a, b) Field dependences of the Hall conductivity $\sigma_{yx}$ and longitudinal conductivity $\sigma_{xx}$ at various temperatures ranging from 10 to 300 K. The black solid lines are fitted curves based on the three-carrier or two-carrier model. (c, d) Temperature dependences of the carrier concentration and mobility extracted from the fitting of the conductivity. The subscripts of e1, e2 denote the electron type, h denote the hole type.

While the multi-carrier model has demonstrated its efficiency in describing transport data, we have supplemented our analysis with first-principles calculations and semiclassical Boltzmann transport theory to further explore the experimental data on CrSb. By taking into account the band details of the material, we aim to provide a more comprehensive understanding of its transport properties.

Since the relaxation time $\tau$ is an uncertain factor in the calculated data [28], we determined the magnitude of $\tau$ (0.12 ps) at 10 K by equalizing the calculated MR to the experimental value at 10 K, 9 T. Then the experimental $\rho_{xx}$ $(T)$, which equals $m^*/ne^2\tau$ $(T)$, is applied to extract other $\tau$ points by assuming constant values of effective mass $(m^*)$ and total carrier concentration $(n)$. The temperature dependence of $\tau$ is shown in the inset of Fig. 5(a). Using these $\tau$ values, we obtain the absolute MR and $\rho_{yx}$ values as a function of magnetic fields, which are plotted in Figs. 5(a, c). Here, the Fermi level has shifted to $E_F$ = 50 meV, indicating a slight electron doping. The corresponding FSs in Fig. 5(b) manifest that the hole pocket narrows compared to that in $E_F$ = 0 meV, and the electron pocket becomes larger and multiply-connected. The calculated MR and $\rho_{yx}$ ($B$) at $E_F$ = 0 eV are shown in the supplementary Fig. S5. It can be seen in both situations the calculated results closely match the key features of the MR variation, with minor deviations at low temperature. However, the trend of $\rho_{yx}$ ($B$) cannot be described by the calculation without considering Fermi level shifting. At $E_F$ = 0 eV, at low temperatures (e.g. 10 K), $R_H$ is completely positive across the whole field region; as the temperature increases, $R_H$ becomes negative at low field, leading to a reversal in sign at a certain field below 9 T. This phenomenon is due to that the critical field $B_{cri}$, where the $R_H$ changes from negative to positive, generally decreases with lowering temperature in a hole-dominated system (detailed discussion can be seen in supplementary Note 2). Similar changes in $B_{cri}$ take place at $E_F$ = 50 meV, but due to the reduced hole population, the sign change persists at lower temperatures. At higher temperatures, $B_{cri}$ exceeds 9 T, making it unobservable in the measured magnetic field region (Fig. 5(c)).

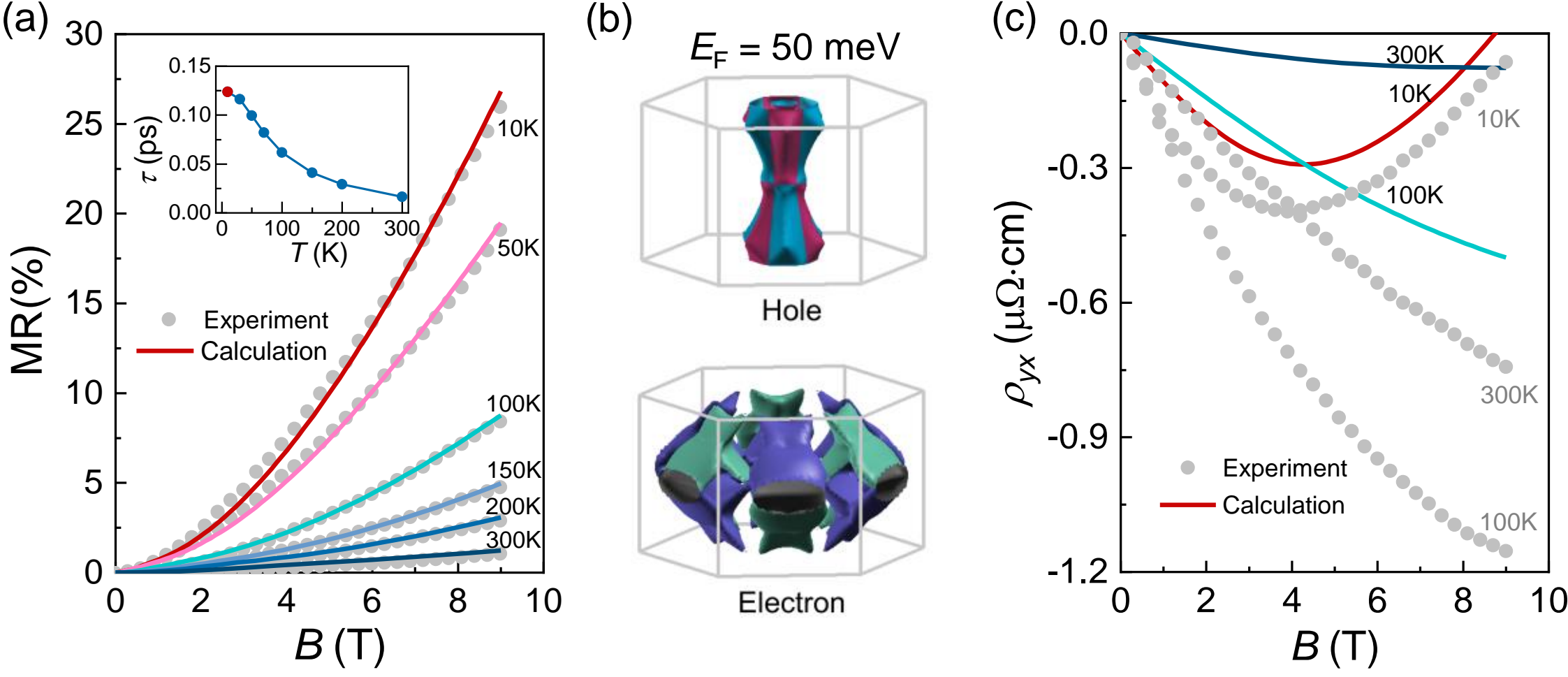

**Fig. 5.** (a, c) Comparison of the calculated MR and Hall resistivity in the electron-doped state ($E_F$ = 50 meV) of CrSb with the measured values at various temperatures ranging from 10 to 300 K. Solid lines represent the calculated values, and gray dots represent the same experimental data shown in Figs. 2(b) and 3(a). The inset in (a) shows the temperature dependence of the relaxation time $\tau$. (b) The spin-resolved Fermi surfaces at $E_F$ = 50 meV.

## 4. Conclusion

In conclusion, we have successfully grown the altermagnetic CrSb in single crystal form and comprehensively investigated its electronic and magneto-transport properties. Large, positive and non-saturated magnetoresistances are identified in CrSb crystals. The behavior of MR well obeys the Kohler's rule, indicating its classic origins. Interestingly, a nonlinear magnetic field dependence of Hall effect that resembles the spontaneous AHE has been identified in large temperature range. After careful analysis of the transport data, we conclude the non-linearity mainly stems from the involvement of different carriers in the magnetoconductivity, rather than from spin-splitting related Berry curvature. Our electronic structure calculations have identified three main types of carriers in CrSb, leading us to apply a three-carrier model to fit the conductivity data, which showed good agreement. The extracted carrier concentration and mobility suggest that CrSb behaves more like a semimetal, with a maximum mobility of $3\times10^3$ $cm^2/Vs$. Furthermore, our calculations using the semiclassical Boltzmann transport theory have successfully reproduced the main features of the experimental data for MR and Hall resistivity. These results indicate that the electrical transport phenomena in CrSb are largely influenced by the specific band structure and Fermi surfaces present in the material. To explore the altermagnetic spin-splitting-related transport properties, techniques such as applying strain or appropriate doping to CrSb may be employed to adjust the Néel vector.

## Acknowledgements

This work is supported by National Natural Science Foundation of China (Grant Nos. 12374114 and 52371189).